\documentclass[conference]{IEEEtran}
\IEEEoverridecommandlockouts
\usepackage{amsmath,amssymb,amsfonts}
\usepackage{url}

\usepackage{graphicx}
\usepackage{float}
\usepackage{multirow}
\usepackage{graphicx}
\usepackage{tcolorbox}
\usepackage{booktabs}
\usepackage{balance}
\def\BibTeX{{\rm B\kern-.05em{\sc i\kern-.025em b}\kern-.08em
    T\kern-.1667em\lower.7ex\hbox{E}\kern-.125emX}}

\newtcolorbox{summarybox}{
  enhanced,
  colback=gray!3,
  colframe=gray!50!black,
  rounded corners,
}

\begin{document}

\title{Exploring Engagement in Hybrid Meetings}

\author{\IEEEauthorblockN{ %
Daniela Grassi\textsuperscript{1}, %
Fabio Calefato\textsuperscript{1}, %
Darja Smite\textsuperscript{2}, %
Nicole Novielli\textsuperscript{1}, %
Filippo Lanubile\textsuperscript{1} %
}

\IEEEauthorblockA{\textsuperscript{1}University of Bari, Bari, Italy, Email: \{daniela.grassi, fabio.calefato, nicole.novielli, filippo.lanubile\}@uniba.it}
\IEEEauthorblockA{\textsuperscript{2}Blekinge Institute of Technology, Karlskrona, Sweden, Email: darja.smite@bth.se}
}

%\IEEEoverridecommandlockouts
%\IEEEpubid{\makebox[\columnwidth]{979-8-3315-9147-2/25/\$31.00~ \copyright2025. IEEE \hfill} \hspace{\columnsep}\makebox[\columnwidth]{ }}

\maketitle

\begin{abstract}
%Background: 
%The shift to hybrid work has introduced new challenges in software development, particularly in communication and collaboration during meetings. Hybrid work, widely adopted following the COVID-19 pandemic, has led to a new norm where even office-first companies embrace hybrid team structures.
%Remote participation has become an everyday occurrence, but it risks leading to isolation, alienation, and decreased engagement among remote team members.
\textit{Background}. The widespread adoption of hybrid work following the COVID-19 pandemic has fundamentally transformed software development practices, introducing new challenges in communication and collaboration as organizations transition from traditional office-based structures to flexible working arrangements. This shift has established a new organizational norm where even traditionally office-first companies now embrace hybrid team structures. While remote participation in meetings has become commonplace in this new environment, it may lead to isolation, alienation, and decreased engagement among remote team members. 
%Objective: 
\textit{Aims}. This study aims to identify and characterize engagement patterns in hybrid meetings through objective measurements, focusing on the differences between co-located and remote participants. 
%Methods: 
\textit{Method}. We studied professionals from three software companies over several weeks, employing a multimodal approach to measure engagement. Data were collected through self-reported questionnaires and physiological measurements using biometric devices during hybrid meetings to understand engagement dynamics. 
%Results
%\textit{Results}. Our findings indicate that remote participants displayed higher engagement than onsite participants, with different patterns observed across meeting durations. Both, remote and onsite participants showed lower engagement in longer meetings, with remote participants exhibiting steeper changes between short and medium-length meetings. Lower engagement levels were observed in larger meetings among onsite participants. Engagement in remote participants varied with body language visibility, whereas participants as listeners showed lower engagement than those in active roles regardless of participation mode.
%The regression analyses revealed no significant difference in work engagement between onsite and remote participants overall. However, remote participants reported steeper declines in engagement during medium-length meetings (30-60 minutes). Work engagement was positively associated with active meeting roles and negatively associated with multitasking behavior. Emotional engagement declined significantly in meetings longer than 30 minutes for all participants. Meeting characteristics such as size and time of day also influenced engagement, with larger meetings and afternoon sessions showing lower engagement levels. Our analysis of physiological synchrony revealed no significant differences in mutual engagement between co-located and distributed participant pairs.
\textit{Results}. The regression analyses revealed comparable engagement levels between onsite and remote participants, though remote participants show lower engagement in long meetings regardless of participation mode. Active roles positively correlate with higher engagement, while larger meetings and afternoon sessions are associated with lower engagement. 
%Conclusions 
\textit{Conclusions}. Our results offer insights into factors associated with engagement and disengagement in hybrid meetings, as well as potential meeting improvement recommendations.
These insights are potentially relevant not only for software teams but also for knowledge-intensive organizations across various sectors facing similar hybrid collaboration challenges.
\end{abstract}


\section{Introduction}
Software companies and other industries employing knowledge workers have undergone significant transformations in recent years, transitioning from predominantly in-office work to fully remote work during the COVID-19 pandemic~\cite{ralph2020pandemic,Ford:etAl:22_pandemic,Butler:Jaffe:ICSE21:pandemic}, and now adapting to a hybrid work reality~\cite{Smite:et:al:2023:WFH,Conboy:etAl:2023:HybridSE}. This shift has led to an unprecedented prevalence of flexplace (allowing employees to choose where to work) and flextime (allowing employees to choose when to work) arrangements. Even companies with office-first policies, requiring some degree of in-office presence, now rely heavily on digitally mediated communication, collaboration, and coordination. Individual choices of working in-office or remotely subsequently impacted the team dynamics, with hybrid team structures becoming the norm. On any given day, it is not uncommon to have team members working together in a co-located office space while others work remotely \cite{Santos2022}. Some team members may even be permanently remote, leading to hybrid and digital meetings being an everyday occurrence.

Hybrid meetings involve co-located participants engaging in coordinated work with remote participants \cite{neumayr2021hybrid, stray2024hybrid}. Although teleworking and distributed work are not new, the concept of hybrid meetings, as they exist today, gained research attention only during the pandemic. In contrast to pre-pandemic distributed meetings, which connected two or more co-located groups in different offices, today's hybrid meetings involve co-located participants collaborating with remote individuals, often working in isolation, typically from home. 

Early studies of hybrid meetings in software companies highlight many challenges. For example, remote participants may feel like ``second-class citizens,'' experiencing alienation, limited opportunities to contribute, difficulty in voicing opinions, disagreeing, or asking questions \cite{tkalich2022}. This imbalance can lead to remote participants taking passive roles or even disengaging from discussions \cite{stray2024hybrid}. Additionally, remote meetings are associated with higher instances of multitasking compared to co-located meetings, likely due to the lower cost of getting noticed \cite{cao2021multitasking, reunamaki2023}. These challenges raise critical questions about whether hybrid working models inadvertently foster disengagement and whether this justifies calls for return-to-office policies. 

Understanding and measuring engagement in hybrid meetings is an important step to address the burning questions related to the future of work. One approach used by Stray et al.~\cite{stray2024hybrid} was to observe meeting participants — how they engage in discussions, who speaks for how long, turn-taking, and participation with (turned on or muted) audio and video (on turned on or off). However, such measurement is rather intrusive and participants may behave differently when feeling observed. Besides, this approach disregards the emotional side of engagement –  the depth of personal investment and emotive connection an employee experiences during an activity such as a meeting \cite{alagaraja2015exploring}. Alternatively, remote participants could be directly asked about their engagement. However, this approach may introduce biases, as those preferring remote work may overreport productivity and engagement to defend their choice. Therefore, objective measures are necessary to gain a clearer picture of engagement and disengagement in hybrid settings. 

In this paper, we propose using physiological response devices to measure engagement and disengagement more objectively. By tracking measurable indicators such % as heart rate variability or
skin conductance, we aim to identify patterns and characteristics of participant engagement in hybrid meetings. Our research is driven by the following question: 

\textbf{RQ}: \textit{What characterizes participants’ engagement in hybrid meetings?}  

To address our research question, we designed and implemented a study protocol leveraging a mixed-methods approach to investigate engagement in hybrid meetings at three Swedish software and telecommunication companies over several weeks. We collected data through two main instruments: a wearable device that captured physiological signals (electrodermal activity) and post-meeting online questionnaires. Thirty software developers joined the study voluntarily, generating 226 meeting-participation data points (70 remote, 156 onsite) and 46 participant dyads for analysis. 
We operationalize engagement along three dimensions, namely \textit{work engagement}, \textit{emotional engagement}, and \textit{mutual engagement}, which we measure using a combination of sensor-based metrics and self-reported engagement scores. 

%The contributions of this work are as follows:
%\begin{itemize}
%    \item We define and validate a multimodal methodology for measuring engagement by combining self-report and wearable devices in workplace settings. 
%    \item We provide initial empirical evidence challenging assumptions about remote participant disadvantages in hybrid meetings. 
%    \item We identify specific factors associated with engagement across different participation modes and provide actionable guidelines to enhance engagement in hybrid meetings.
%    \item We build and distribute a replication package to encourage future replication of the present study.\footnote{https://doi.org/10.6084/m9.figshare.28323638.v1}\fabio{in realtà il dataset non lo diamo piu', quindi non è tanto un replication package, quanto gli analysis script}
%\end{itemize}

The contributions of this work are multifaceted. We define and validate a multimodal methodology for measuring engagement by combining self-report and wearable devices. Furthermore, we provide initial empirical evidence that challenges common assumptions about remote participant disadvantages in hybrid meetings. Through our analysis, we identify specific factors associated with engagement across different participation modes, along with actionable guidelines that can enhance engagement in hybrid meetings. 
To support research reproducibility, we make our analysis scripts available through a public repository,\footnote{https://figshare.com/s/85a3c7e7a29db0006dd2} enabling other researchers to apply similar analytical approaches. We do not distribute the data from our study as per the agreement with the companies involved.

The remainder of this paper is organized as follows. 
Section~\ref{sec:background} establishes the theoretical foundation, by examining the dimensions of engagement, and reviews existing approaches for measuring these constructs. 
Section~\ref{sec:studydesign} presents our research methodology, including our conceptual framework for assessing engagement in hybrid meetings, the characteristics of participating software companies, and our multi-modal data collection approach combining physiological measurements with self-reported data. 
Section~\ref{sec:results} reports our findings, which we discuss in Section~\ref{sec:discussion}, along with a comparison with existing literature on hybrid meetings and a discussion of limitations. We provide concluding remarks in Section~\ref{sec:conclusion}. 

\section{Background}
\label{sec:background}

In this section, we present the theoretical foundations of engagement and its measurement. Specifically, we first define the three dimensions of engagement we investigate in this study. Then, we review the various methods used in literature for measuring engagement, from self-report to physiological measures, explaining how they enable capturing different aspects of engagement in meetings and interactions.

\subsection{Engagement}
Engagement is often perceived simply as active participation, such as speaking up and gesticulating \cite{stray2024hybrid}. However, the theoretical concept of engagement is far more complex, encompassing not only behavioral but also emotional and cognitive dimensions that interact dynamically within various contexts. \textit{Engagement} is defined as a dynamic process through which two or more participants establish, maintain, and conclude a perceived connection~\cite{kahn_psychological_1990}. This process is fundamental to meeting success, as it directly influences the quality of communication, productivity, and effectiveness of group work~\cite{said_exploring_2020}. Our research focuses on three specific yet complementary dimensions of engagement: work, emotional, and mutual engagement.

\textit{Work engagement}, as defined by Schaufeli et al.~\cite{schaufeli2003utrecht}, represents a positive, fulfilling, work-related state of mind that manifests through three distinct but interconnected components \cite{schaufeli2002measurement}. 
The first component, \textit{vigor}, is characterized by high energy levels and mental resilience while working, enabling employees to persist through challenges and invest sustained effort in their tasks. 
The second component, \textit{dedication}, reflects an employee's deep psychological involvement in their work, encompassing feelings of significance, enthusiasm, and challenge that drive meaningful contributions. 
The third component, \textit{absorption}, describes an employee's state of full concentration and immersion in their work, leading to enhanced focus and productivity. 
These three components work in concert to produce significant organizational benefits: vigorous employees demonstrate higher levels of personal initiative~\cite{salanova2008cross}, dedicated workers show increased organizational commitment~\cite{hakanen2008job}, and absorbed employees exhibit improved performance outcomes~\cite{bakker2004using}. 
Research has further established that work engagement correlates positively with employee health outcomes~\cite{schaufeli2008workaholism} and work-unit innovativeness~\cite{hakanen2008positive}, while negatively associating with turnover intention \cite{schaufeli2004job} and sickness absence~\cite{schaufeli2009changes}. 
This comprehensive impact on individual and organizational outcomes underscores work engagement's crucial role in workplace well-being and productivity.

\textit{Emotional engagement} represents the depth of personal investment and emotive connection an employee experiences with their work activities \cite{alagaraja2015exploring}. 
This dimension emerges when employees develop a personal connection with organizational goals and identify with their work on an affective level. 
Employees translate this emotional investment into action by contributing their knowledge, skills, and abilities toward task achievement with an intense focus~\cite{alagaraja2015exploring}.
This dimension is particularly significant in hybrid work environments, where maintaining emotional connections presents unique challenges \cite{tkalich_what_2022,miller2021how}. 
In hybrid meetings, remote participants often become passive observers, with the co-located group's activity dominating discussions \cite{tkalich_what_2022, stray2024hybrid}. 
Technical delays and the absence of non-verbal cues further discourage remote participants from engaging actively \cite{OLearyM10, stray2024hybrid}, leading them to remain primarily in listening mode. 
This lack of non-verbal communication also impacts emotional expression, making it more difficult for participants to express and perceive emotions in virtual interactions \cite{miller2021how}.

\textit{Mutual engagement} occurs when two or more individuals experience aligned levels of self-reported engagement during shared activities~\cite{gashi_using,NGUYEN2021118599:mutual:engagement,Piazza:mutual:engagement}.
This alignment of engagement states is critical for creative collaborations and team-based activities, as it goes beyond mere simultaneous participation~\cite{identifying_bryankinns_2012}. 
Research has demonstrated that mutual engagement manifests through physiological synchrony between participants, that is, the association and interdependence between their physiological states during interactions~\cite{gashi_using}.
However, achieving an alignment in engagement states is distinctly challenging for hybrid meeting participants.
Unlike face-to-face interactions, where non-verbal cues and immediate feedback naturally facilitate alignment, hybrid settings introduce communication asymmetries between participants~\cite{identifying_bryankinns_2012}. 
Limited access to social cues for remote participants can impair mutual engagement between those in physical and virtual spaces.
Understanding mutual engagement in hybrid meetings can provide valuable insights and inform the design of practices that promote higher levels of engagement for remote and onsite participants~\cite{kuzminykh_classification_2020}. This is becoming increasingly critical as hybrid work models gain relevance across organizations.

\subsection{Measuring Engagement}
Researchers have employed diverse approaches to measure human engagement across different contexts and activities. 
Self-report measures provide direct access to subjective experiences, making them a popular assessment tool, although participants may struggle to recall past experiences accurately~\cite{monkaresi2016automated}.
This has motivated the exploration of less intrusive physiological measures. 

Electroencephalography (EEG) has shown promise in detecting engagement levels during meetings~\cite{microsoft} and learning activities~\cite{apicella2022eeg}, though its practical application remains limited by the required equipment. 
More recent work has focused on less invasive measures like facial expressions, heart rate~\cite{monkaresi2016automated}, and electrodermal activity (EDA) captured through wearable devices.
EDA, also known as galvanic skin response, measures changes in skin conductance and indicates physiological arousal closely tied to attention and engagement~\cite{hernandez2014using}. 
Di Lascio et al.~\cite{dilascio2018} developed novel features to detect \textit{momentary engagement}, i.e., periods when subjects show a heightened interest and readily engage in the current activities. 
EDA signals are processed in several steps. 
The first step involves discretizing the continuous EDA signal into five distinct levels, which capture different intensities of physiological arousal.
Building on this discretization, Di Lascio et al. analyze transitions between these levels to identify \textit{arousing moments}, which they define as time intervals where the EDA level shows an increase. 
To quantify these dynamics over time, they introduce the \textit{arousing ratio} feature that captures the relative frequency of such moments by calculating the ratio between arousing and unarousing moments during a session.
Their findings reveal that the arousing ratio and the percentage of time spent at the highest EDA level (level 5) are the two most effective features in discriminating between engaged and non-engaged subjects.

Beyond individual measures, researchers have increasingly examined physiological synchrony between participants as an indicator of mutual engagement. 
This synchrony, defined as the temporal association between physiological states of interacting individuals \cite{karvonen2016sympathetic}, has proven informative across various contexts, including child-adult interactions \cite{hernandez2014using}, therapeutic relationships \cite{marci2007physiologic}, and audience-presenter dynamics \cite{gashi_using}. 
Slovák et al.~\cite{slovak2014exploring} found that increased EDA signal synchrony between friends correlates with heightened emotional engagement in social interactions.
Gashi et al.~\cite{gashi_using} investigated the potential of using EDA as a physiological measure to assess mutual engagement between presenters and their audience. 
After evaluating seven distinct algorithmic approaches, they found that Dynamic Time Warping (DTW)~\cite{muller2007dynamic} provided the most reliable method for quantifying engagement synchronization by capturing the dynamic nature of engagement patterns during presentations.
In particular, DTW identifies patterns in time-series data by finding optimal ways to match two sequences, even when they unfold at different rates. By stretching or compressing the timing of one sequence to best match the other, DTW can detect similar patterns despite variations in their temporal dynamics. 
This flexible approach makes DTW particularly effective for comparing physiological signals that naturally vary in timing across different individuals.

\section{Study Design}
\label{sec:studydesign}

This section presents the research methodology used to investigate engagement in hybrid meetings.

\subsection{Conceptual Framework} 
\begin{figure}[b]
    \centering
    \vspace{-4.5mm}
    \includegraphics[width=0.60\linewidth]{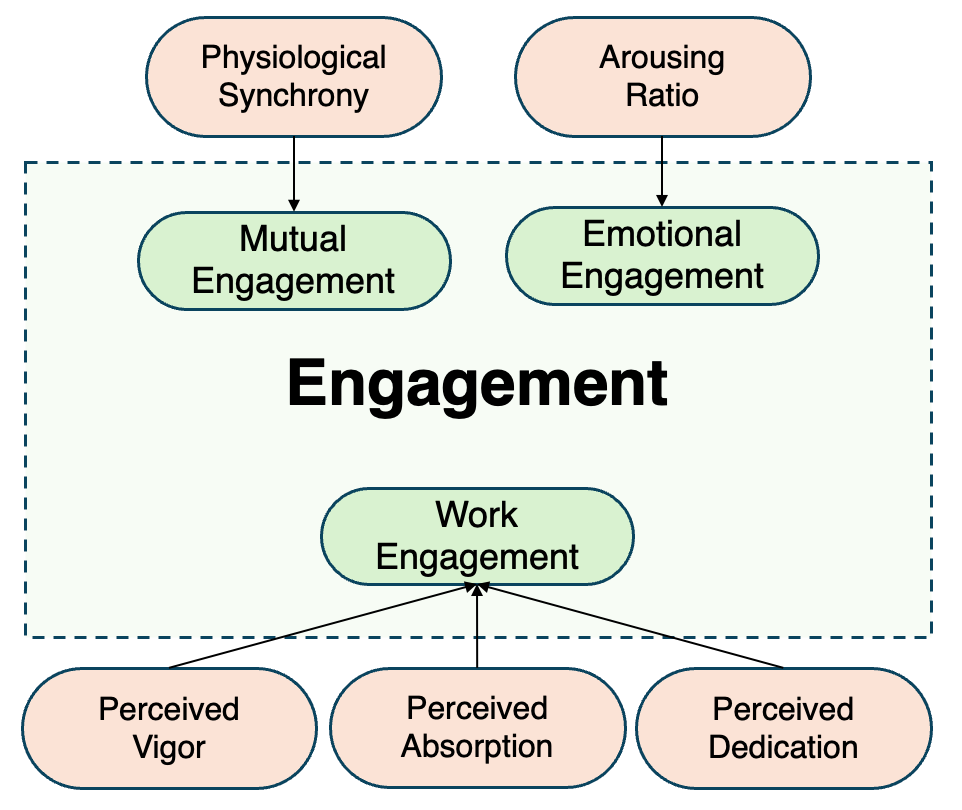}
    \caption{Conceptual framework for assessing the three dimensions of engagement in hybrid meetings.}
    \label{fig:framework}
\end{figure}

In the following, we present a conceptual framework for the analysis of factors influencing participants' engagement in hybrid meetings. 
The framework, illustrated in Figure~\ref{fig:framework}, is built upon previous studies~\cite{dilascio2018, gao_n-gage_2020, gashi_using, ferrari2024using} investigating the three dimensions of engagement -- work, emotional, and mutual -- presented in Section~\ref{sec:background}.
Also, we derive metrics from the literature to assess each of the three factors.
In particular, for emotional and mutual engagement, we collected physiological signals, whereas for work engagement, we collected self-reported measures through a survey administered to participants at the end of meetings.

\subsubsection*{Work Engagement} 
\label{sec:framework_work}

Given the absence of validated physiological metrics for assessing work engagement in the existing literature, we employed a self-report approach using an adapted version of the Utrecht Work Engagement Scale (UWES-9) \cite{schaufeli2003utrecht}, an instrument previously validated in meeting-focused studies~\cite{lehmann-willenbrock_our_2016, nurmi_virtual_2023}. 
Although UWES-9 comprises multiple items across three subscales--vigor, dedication, and absorption--its standard items address general work scenarios (e.g.,``\textit{I can continue working for very long periods at a time},'' ``\textit{I am enthusiastic about my job}'') rather than specific meeting contexts. 
Therefore, we adapted one item per subscale to better reflect meeting participation dynamics.
The adapted items -- see the first three rows of Table~\ref{tab:questions} -- maintained the core constructs while focusing specifically on meeting experiences: vigor was measured by asking respondents to evaluate the statement ``\textit{During the meeting, I felt full of energy}''; for dedication we used the statement – ``\textit{I found this meeting full of meaning and purpose}''; and for absorption we used the statement – ``\textit{During this meeting, time flew}.'' 
For each item, participants reported an engagement score on a 5-point scale. 
Following the same approach used by Nurmi and Pakarinen~\cite{nurmi_virtual_2023}, we calculated the overall work engagement score as the arithmetic mean of these three components.

\subsubsection*{Emotional Engagement}
To quantify participants' emotional engagement during hybrid meetings, we adopted the \textit{arousing ratio} metric derived from EDA signals, following the methodology established by Di Lascio et al.~\cite{dilascio2018}. 
Their approach involves a systematic process of signal discretization and analysis to identify periods of heightened physiological arousal.
The method applies piecewise aggregate approximation (PAA) to discretize an EDA signal into five levels of equal width. Through PAA, the signal is first segmented into blocks, with mean values computed for each block to generate a sequence $S$. 
This sequence is then mapped to five equidistant levels, producing a discretized signal $S_l$. 
The relative change between consecutive levels is calculated as $\Delta l = S_l[i] - S_l[i-1]$, where $S_l[i]$ represents the current level.
Using 30-second time intervals as the measurement period, we classified signal segments as either arousing or unarousing moments. 
Intervals where $\Delta l \geq 1$ were designated as arousing moments, indicating an increase of at least one discrete level. 
All other intervals ($\Delta l < 1$) were classified as unarousing moments. 
From the available metrics derived from this classification, we selected the arousing ratio--the proportion of arousing to unarousing moments--due to its signal length independence, making it particularly suitable for comparing meetings of varying durations.

\subsubsection*{Mutual Engagement}
Prior research has demonstrated that physiological synchrony between EDA signals serves as an effective proxy for measuring mutual engagement between participants~\cite{gashi_using}.
Their findings identified the Dynamic Time Warping (DTW) algorithm as the most reliable method for determining optimal alignment between time series. 
Building upon this work, we chose FastDTW,\footnote{https://pypi.org/project/fastdtw} a Python-based implementation that offers enhanced control over locality constraints to regulate signal compression and dilation~\cite{muller2007dynamic}. 
We utilized a locality constraint of one EDA sample, a parameter previously validated for measuring mutual engagement~\cite{gashi_using}.

\subsubsection*{Meeting and Participation Characteristics}
Beyond measuring the three dimensions of engagement and how they vary with participation mode, we also consider other non-actionable factors associated with the meeting organization and individual participation. 
Specifically, we investigate two categories of contextual factors. 

The \textit{meeting characteristics} encompass inherent aspects of the meetings that are typically determined by team and company needs rather than individual choice. 
These include the meeting's duration, categorized as shorter (up to 30 minutes), mid-length (30 to 60 minutes), or extended (over 60 minutes) based on questionnaire responses. Similarly, we classified meeting size into smaller sessions (2-4 participants) or larger sessions (more than 5 participants). We considered the type of meeting—whether brainstorming, customer, stand-up, or other formats. Additionally, we examined temporal factors by grouping meetings by day of the week (Monday through Friday) and time of day, with the latter classified into five distinct windows: Early Morning (5–10 AM), Before Lunch (10 AM–12 PM), Lunch Time (12–1 PM), After Lunch (1–3 PM), and Afternoon (3–8 PM).

The \textit{participation characteristics} represent individual-level factors that could impact engagement while being constrained by practical considerations. They include participants' role (e.g., listener, presenter/speaker, discussant, leader), their participation mode (onsite or remote), self-reported multitasking behavior (High or Low), and their daily meeting load (i.e., one meeting per day, two, or more than two). 

%To ensure model integrity and avoid potential bias, we excluded factors related to the visibility of facial expressions and body language due to their strong mutual correlation (Cramer's V = 0.88). Furthermore, these variables were strongly imbalanced differently from the others. %distribution in our sample. The data revealed a significant disparity: while 181 participants reported full visibility of facial expressions, only 45 reported partial visibility. Similarly, for body language, 188 participants reported full visibility compared to only 38 with partial visibility. 

%Although it is not always possible to control these characteristics in real-world settings, e.g. due to specific or contingent organizational needs, understanding their potential influence on engagement in meetings is crucial for comprehensive analysis.

\subsection{Companies and Participants}
Our study examined hybrid meetings in real-life settings across three Swedish software companies spanning different sizes and domains. To maintain anonymity, we refer to these companies using fictional names.
The first company, AlphaConsult, is a small consulting firm delivering bespoke software development services to its clients.
The second company, BetaSolutions, is a medium-sized Swedish branch of a large enterprise delivering diverse software-intensive products.
The third company, GammaCorp, is a large tech company operating in the telecom industry. 
These organizations were selected based on their expressed commitment to evaluating and enhancing their meeting culture, with management actively supporting the research initiative. 
Following discussions with key stakeholders at each company, we established a recruitment process that emphasized voluntary participation. 

To ensure appropriate data collection and success with our study, we collaborated with management to identify and recruit teams that regularly conducted hybrid meetings as part of their standard work practices.
In AlphaConsult, the manager invited two teams to participate in the study. Both teams included software developers, team leads, and project managers. One of the teams had three permanently remote team members. Besides, members of both teams had regular hybrid meetings with the customers. 
At BetaSolutions, we distributed the study announcement company-wide. In response, a manager leading an internal open innovation initiative volunteered her team to participate in the study because hybrid meetings were integral to their regular workflow.
In GammaCorp, several managers attended our presentation about hybrid meetings and expressed interest in participating in the study. %All managers belonged to the same corporate forum and had many hybrid meetings in their schedules. 

To facilitate the recruitment of individual participants, two authors conducted separate briefing meetings with each company to outline the study's objectives, the requirements for participants' involvement, the experimental protocol, and the potential risks and benefits of participation. Upon verbal instructions, potential participants received written informed consent forms to sign and instruments (see details in the next section). Signing the forms was voluntary, and participants at any moment could withdraw from participation. 

In total, the study involved 30 participants, 22 from AlphaConsult, 6 from BetaSolutions, and 2 from GammaCorp. To preserve participants' privacy and confidentiality, all data were fully anonymized, and demographic data were not collected.

\subsection{Instrumentation}

To collect data, we employed two main instruments: a wearable device to collect physiological signals and a post-meeting questionnaire to gather self-reported engagement measures and contextual information.

The Embrace Plus wristband\footnote{https://www.empatica.com/en-eu/embraceplus/} served as our primary tool for capturing physiological data relevant to the emotional and mutual engagement dimensions. 
This device is equipped with an EDA sensor enabling the measurement of physiological arousal and synchrony between participants, with a sampling frequency of 4Hz. 
We requested participants to wear the wristband throughout their workday to avoid the need to spend time switching it on and off, which might influence the data.

To complement the physiological measurements, participants were instructed to complete a web-based questionnaire, shown in Table~\ref{tab:questions}, after each meeting.
To facilitate convenient access to the questionnaire, QR codes linking to the online questionnaire form were placed in participants' offices.
Beyond work engagement metrics, a questionnaire was also designed to capture the meeting and participation characteristics (already presented in Section~\ref{sec:framework_work}). The possible values for these characteristics were defined in agreement with the companies participating in the study.

\begin{table*}[tb]
\caption{Post-meeting questionnaire to assess work engagement and collect information about meeting and participation characteristics.}
\centering
\resizebox{\textwidth}{!}{%
\begin{tabular}{c|l|l}
\hline
 & \multicolumn{1}{c|}{\textbf{Questions}} & \multicolumn{1}{c}{\textbf{Answer Type}} \\ \hline
\multirow{3}{*}{\begin{tabular}[c]{@{}c@{}}\textbf{Work} \\ \textbf{Engagement}\end{tabular}} & During the meeting, I felt full of energy & \multirow{3}{*}{5-point scale with scores in [1-5]} \\
 & During the meeting, time flew &  \\
 & I found this meeting full of meaning and purpose &  \\ \hline
\multirow{4}{*}{\begin{tabular}[c]{@{}c@{}}\textbf{Meeting}\\ \textbf{Characteristics}\end{tabular}} & Starting time of the meeting & Date/time \\
% & Planned duration of the meeting & In minutes \\
 & Duration of the meeting & In minutes \\
 & Meeting size & Number of participants \\
 & Type of meeting & Categorical (brainstorming, customer, stand-up, status, other) \\ \hline
\multirow{6}{*}{\begin{tabular}[c]{@{}c@{}}\textbf{Participation} \\ \textbf{Characteristics}\end{tabular}} & Meeting location & Categorical (onsite or remote) \\
 & Role in the meeting & Categorical (leader, presenter/speaker, discussant, listener, other) \\
 & Multitasking during the meeting & Ordinal (never, rarely, sometimes, often, all the time) \\
 & Ability to see facial expressions of other participants & Categorical (none, some participants, only onsite, only remote, all participants) \\
 & Ability to see body language of other participants & Categorical (none, some participants, only onsite, only remote, all participants) \\
 & Keeping the camera on during the meeting & Ordinal (never, rarely, sometimes, often, all the time) \\ \hline
\end{tabular}%
}
\vspace{-4mm}
\label{tab:questions}
\end{table*}

\subsection{Study Execution}

The study execution consisted of two main phases: preparation and data collection. 
During the preparation phase, two authors conducted a briefing meeting with study participants to demonstrate proper usage of the Embrace Plus wristband and the online questionnaire form for meeting data collection. 
After addressing participants' questions, informed consent forms were signed. 
For remote participants, wristbands were shipped to their homes along with detailed setup instructions and consent forms.

\begin{figure}[b]
\centering
\includegraphics[width=0.4\textwidth]{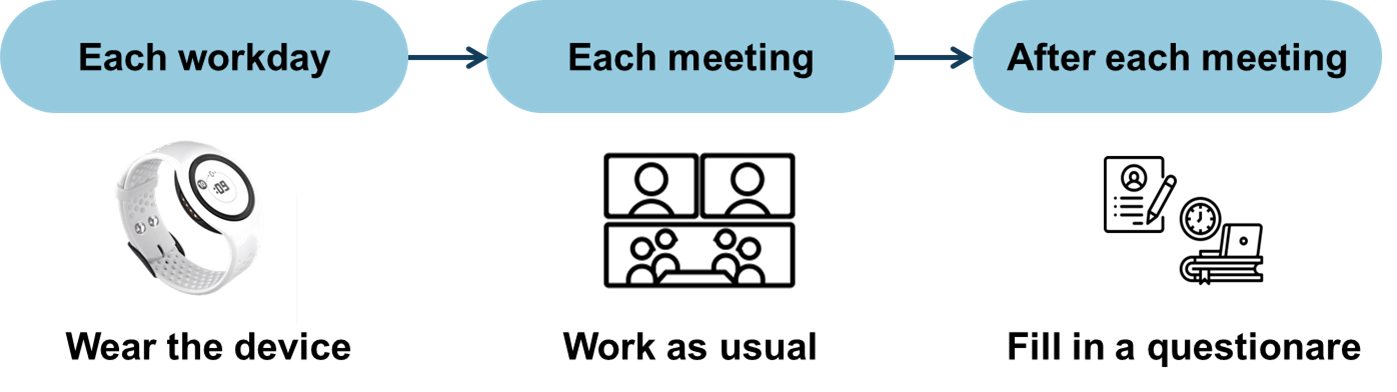}
\caption{An overview of participants' involvement.}
\vspace{-3mm}
\label{fig:timeline}
\end{figure}

The data collection phase began the following day and continued for one to three weeks, depending on both participants' availability and the sufficiency of data collected. 
In particular, one team in AlphaConsult and the team from BetaSolutions participated in the study for one week each, while the second team from AlphaConsult and the two managers from GammaCorp participated for two weeks. %because of the low number of meetings during the first week of their participation in the study.  
Participants wore the Embrace Plus wristband throughout their workday and completed a post-meeting questionnaire after each meeting. Figure~\ref{fig:timeline} depicts a short overview of the participants' involvement in the study. 
The wristband automatically synchronized and uploaded biometric data to the Empatica research portal, from which we downloaded raw data. %in CSV and AVRO formats.
The first author performed weekly data reviews to ensure consistency and completeness, with email follow-ups when clarification was needed.

%\nicole{Check the data in the following paragraph if we include the new data points.} 
Upon study completion, we conducted a final meeting where we presented a company-level overview of participants' engagement experiences.
Overall, our analysis included 228 meeting records with biometric data and complete online questionnaire responses. Two records were excluded due to the biometric device malfunctioning, and 5 records were excluded because the participants did not respond to the post-meeting survey.
Therefore, the final dataset comprised 70 remote and 156 onsite records of meeting participation for the analysis of emotional engagement, and 153 onsite and 70 remote for the analysis of work engagement.
To analyze mutual engagement states, i.e., physiological synchrony, between meeting participants, we identified 46 dyads of participants who attended the same meetings. 
These dyads included 36 co-located pairs (where both participants were onsite) and 10 distributed pairs (where at least one participant was remote).

\subsection{Analysis}
To assess engagement along its three dimensions, we analyzed both physiological measurements and self-reported data.

Before building our statistical models, we examined correlations between all predictors to identify potential multicollinearity issues. For categorical variables, we used Cramer's V coefficient to measure the strength of association. Notably, we found a strong correlation between the visibility of facial expressions and body language (Cramer's V = 0.88), indicating these variables essentially captured the same underlying construct. To ensure model integrity and avoid potential bias, we excluded both factors (visibility of facial expressions and body language) from the subsequent analyses because they also showed strongly imbalanced distributions compared to other predictors, which could compromise the model stability.

To assess how participation mode and contextual meeting factors associate with the three engagement dimensions, we built different regression models.
For work and emotional engagement analysis, we fitted two fixed effects (FE) models to account for stable differences between companies and individuals. Since work and emotional engagement scores are influenced by constant factors like company culture and participants' physiology, FE models can effectively control for these unobserved characteristics without requiring their direct measurement. By removing these potential confounding variables, the fixed effects approach allowed us to more precisely examine how specific meeting and participant characteristics of interest are associated with engagement.
%For emotional engagement, we built a generalized linear mixed-effects model (GLMM)  with a Gamma distribution and log link function due to the right-skewed distribution of the arousing ratio values. This modeling approach accommodated the non-normal distribution of our data while enabling robust parameter estimation. Both models incorporated fixed effects to account for actionable predictors—such %as meeting type, duration, day of the week, meeting size, multitasking, and time of day—while controlling for non-actionable factors (e.g., company culture) and intrinsic individual differences of participants.
We also included interaction terms between participation mode and other factors (e.g., multitasking, meeting type and duration) to evaluate conditional effects that might emerge only in specific contexts.
%To analyze work engagement as the dependent variable, we built a fixed effects model, as its distribution of values are bounded in the interval [1,5]. To analyze emotional engagement, we fitted a generalized linear mixed-effects models (GLMM) with a Gamma distribution and log link function, given the right skewed distribution. These models incorporated fixed effects to account for actionable predictors—such as meeting type, duration, day of the week, meeting size, multitasking, and time of day—as well as control variable for non-actionable factors (e.g., company culture) and intrinsic individual differences of subjects. We also analyzed the interaction terms between participation mode and other factors (e.g., multitasking, meeting type, duration) to evaluate conditional effects. 

Regarding mutual engagement, the limited number of dyads identified precluded the use of complex models. We therefore fitted a pooled Ordinary Least Squares (OLS) model, retaining only participation mode (our primary focus) and meeting duration (due to the time-dependent nature of physiological synchrony measures), with observations clustered by dyads. To address the non-independence of observations within the same dyad, we implemented clustered standard errors to account for potential heteroscedasticity and intra-cluster error correlation. 
Despite these limitations, this approach was sufficiently robust and allowed us to perform an exploratory analysis of potential differences in mutual engagement between co-located and distributed dyads while controlling for meeting duration.

\section{Results}
\label{sec:results}

In this section, we present the results of our analysis of engagement in hybrid meetings across three dimensions: work engagement based on self-reported data (Section~\ref{sec:res_work_eng}), emotional engagement measured through arousing ratio (Section~\ref{sec:res_emo_eng}), and mutual engagement assessed through physiological synchrony between participants (Section~\ref{sec:res_mutual_eng}).

\subsection{Changes in Work Engagement}
\label{sec:res_work_eng}

The results of the regression analysis for work engagement are presented in Table~\ref{tab:fixed_effects_work}. Contrary to common assumptions about remote participation, our analysis revealed no statistically significant main effect of participation mode on work engagement, indicating that, overall, remote participants reported work engagement levels similar to their onsite counterparts.

\begin{table}[t]
    \centering
    \vspace{-1mm}
    \caption{Coefficient estimates for work engagement (default variables are indicated in parentheses).}
    \resizebox{\columnwidth}{!}{
    \begin{tabular}{lccccc}
        \toprule
        \textbf{Coefficient} & $\boldsymbol{\beta}$  & $\boldsymbol{SE}$ & $\boldsymbol{t}$ & $\boldsymbol{p}$ & $\boldsymbol{95\% \ CI}$ \\
        \midrule
 
     \textbf{Participation (Onsite)} \\
        \quad Remote & 0.02 & 0.19 & 0.12 & 0.908 & [-0.34, 0.39] \\
        \textbf{Meeting Type (Status Meeting)} \\
        \quad Brainstorming Meetings & 0.08 & 0.05 & 1.51 & 0.133 & [-0.02, 0.18] \\
        \quad Customer Meetings & -0.07 & 0.06 & -1.17 & 0.243 & [-0.20, 0.05] \\
        \quad Others & 0.09 & 0.05 & 1.61 & 0.109 & [-0.02, 0.20] \\
        \quad Stand-up Meetings & -0.01 & 0.05 & -0.12 & 0.907 & [-0.11, 0.09] \\
        \textbf{Duration (Up to 30 min)} \\
        \quad 30-60 min & -0.01 & 0.04 & -0.19 & 0.851 & [-0.09, 0.07] \\
        \quad $>$60 min & -0.06 & 0.06 & -0.90 & 0.369 & [-0.18, 0.07] \\
        \textbf{Day of Week (Monday)} \\
        \quad Tuesday & -0.04 & 0.05 & -0.80 & 0.426 & [-0.14, 0.06] \\
        \quad Wednesday & -0.01 & 0.06 & -0.17 & 0.865 & [-0.13, 0.11] \\
        \quad Thursday & -0.01 & 0.06 & -0.24 & 0.814 & [-0.12, 0.10] \\
        \quad Friday & -0.01 & 0.07 & -0.17 & 0.863 & [-0.14, 0.12] \\
        \textbf{Meeting Size (2-4)} \\
        \quad 5+ (**) & -0.11 & 0.04 & -2.71 & 0.007 & [-0.19, -0.03] \\
        \textbf{Role (Listener)} \\
        \quad Discussant (*) & 0.14 & 0.07 & 2.01 & 0.046 & [0.00, 0.29] \\
        \quad Leader & 0.11 & 0.09 & 1.20 & 0.231 & [-0.07, 0.30] \\
        \quad Presenter/Speaker (*) & 0.20 & 0.09 & 2.29 & 0.023 & [0.03, 0.36] \\
        \textbf{Multitask (Low)} \\
        \quad High & -0.12 & 0.06 & -1.94 & 0.054 & [-0.24, 0.00] \\
        \textbf{Time of Day (Early Morning)} \\
        \quad Before Lunch & -0.04 & 0.05 & -0.84 & 0.401 & [-0.14, 0.05] \\
        \quad Lunch Time & -0.07 & 0.08 & -0.87 & 0.384 & [-0.23, 0.09] \\
        \quad After Lunch & -0.07 & 0.05 & -1.46 & 0.146 & [-0.16, 0.02] \\
        \quad Afternoon (*) & -0.22 & 0.08 & -2.57 & 0.011 & [-0.38, -0.05] \\
        \textbf{Cumulative Meetings per Day (1)} \\
        \quad 2 meetings & 0.02 & 0.04 & 0.57 & 0.569 & [-0.06, 0.11] \\
        \quad $>$2 meetings & -0.03 & 0.06 & -0.50 & 0.621 & [-0.14, 0.08] \\
        \textbf{Interaction Effects} \\
        \quad Remote × 30-60 min (*) & -0.16 & 0.08 & -2.15 & 0.033 & [-0.31, -0.01] \\
       
        \midrule
        \multicolumn{5}{l}{$R^2$: 0.31 \quad Adj. $R^2$: 0.14 \quad F-statistic: 1.92 (p = 0.002)} \\
        \bottomrule
        \multicolumn{5}{l}{Significance codes: '***' 0.001 '**' 0.01 '*' 0.05} \\
        \end{tabular}
        }
        \label{tab:fixed_effects_work}
        \vspace{-5mm}
    \end{table}

We found that some meeting characteristics are significantly associated with work engagement. Specifically, we observe that participants in larger meetings (5+ participants) reported $11\%$ lower engagement than those involved in smaller meetings ($\beta = -0.11$, $p = 0.007$). 
The time of meetings is also significantly associated with work engagement, with participants in afternoon meetings reporting $22\%$ lower engagement ($\beta = -0.22$, $p = 0.011$) than participants in morning meetings. 

Work engagement showed significant associations also with participant roles. Compared to listeners, both discussants and presenters/speakers reported higher work engagement, respectively $14\%$ ($\beta = 0.14$, $p = 0.046$) and $22\%$ ($\beta = 0.020$, $p = 0.02$).

While the main effect of participation mode was not significant, we found a meaningful interaction with meeting duration. 
Remote participants reported $16\%$ lower work engagement during medium-length meetings (30-60 minutes) than their onsite counterparts ($\beta = -0.16$, $p = 0.033$).
The lack of significant differences for remote participants in meetings longer than one hour is likely due to the smaller number of data points in this category ($n=27$, $12\%$), which may have reduced the statistical power for detecting potential effects.
%This suggests that while remote participation does not inherently reduce engagement, extended virtual meetings may pose particular challenges for maintaining engagement among remote participants. (spostato nella discussione)

%We examined differences in work engagement between onsite and remote participation using our GLMM, results are shown in table \ref{tab:fixed_effects_work}. The analysis revealed no statistically significant main effect of participation mode.

%Additional factors influencing work engagement included participant role and multitasking behavior. Discussants ($\exp(\beta) = 1.13$, $p = 0.02$) and presenters/speakers ($\exp(\beta) = 1.15$, $p = 0.02$) exhibited 13\% and 15\% higher work engagement respectively compared to listeners, while high multitasking was associated with 10\% reduced engagement ($\exp(\beta) = 0.90$, $p = 0.02$).

%Furthermore, meetings held in the afternoon ($\exp(\beta) = 0.86$, $p = 0.01$) were associated with 14\% lower work engagement, and days with multiple meetings (more than two) ($\exp(\beta) = 0.89$, $p = 0.01$) were associated with 11\% reduced work engagement.

%We found a significant interaction between participation mode and meeting duration. Remote participants showed a more pronounced decrease in work engagement during medium-length meetings (30-60 minutes) compared to their onsite counterparts ($\exp(\beta) = 0.86$, $p = 0.01$), with engagement levels 14\% lower in this condition.

\begin{tcolorbox}[standard jigsaw,
title=Key Takeaways - Work Engagement, opacityback=0]
Our analysis revealed no significant difference in \textbf{work engagement} between onsite and remote participants overall. However, remote participants reported lower work engagement during medium-length meetings (30–60 minutes).
Work engagement was positively associated with active meeting roles (discussants, presenters/speakers). Among the meeting characteristics, size and time also correlate with engagement levels, with larger meetings (5+ participants) and afternoon sessions associated with lower engagement.
\end{tcolorbox}

\subsection{Changes in Emotional Engagement}
\label{sec:res_emo_eng}

%We examined factors affecting emotional engagement in meetings using a fixed effects model. Table \ref{tab:fixed_effects_emo} presents the $\beta$ coefficient, ... standard errors, p-values, and 95\% confidence intervals for all fixed effects.

%Our analysis of emotional engagement (see Table~\ref{tab:fixed_effects_emo}) revealed significant effects of meeting type on emotional engagement. Specifically, brainstorming meetings exhibited significantly lower arousing ratio compared to status meetings ($\exp(\beta) = 0.66$, $p = 0.04$), indicating a 34\% decrease in emotional engagement.

The results of our analysis of emotional engagement are reported in Table~\ref{tab:fixed_effects_emo}. As already observed for work engagement, participation mode does not significantly correlate with emotional engagement. 
Among meeting characteristics, meeting duration exhibited a negative correlation with emotional engagement, with a significant $72\%$ decrease for meetings exceeding 60 minutes compared to short meetings (up to 30 minutes) ($\beta = -0.72$, $p = 0.016$).
Moreover, our analysis revealed a significant interaction effect between participation mode and meeting type, whereby remote participants in brainstorming meetings experiencing $84\%$ higher emotional engagement than their onsite counterparts ($\beta = 0.84$, $p = 0.038$).

\begin{tcolorbox}[standard jigsaw,
    title=Key Takeaways - Emotional Engagement, opacityback=0]
    Our analysis revealed that \textbf{emotional engagement} was significantly lower in meetings longer than 60 minutes. We also found that, during brainstorming meetings, emotional engagement was significantly higher for remote participants. 
\end{tcolorbox}

\begin{table}[h]
    \centering
    \caption{Coefficient estimates for emotional engagement (default variables are indicated in parentheses).}
    \resizebox{\columnwidth}{!}{
    \begin{tabular}{lccccc}
        \toprule
        \textbf{Coefficient} & $\boldsymbol{\beta}$ & $\boldsymbol{SE}$ & $\boldsymbol{t}$ & $\boldsymbol{p}$ & $\boldsymbol{95\% \ CI}$ \\
        \midrule
 
         \textbf{Participation (Onsite)} \\
            \quad Remote & -1.22 & 0.85 & -1.43 & 0.156 & [-0.94, 0.00] \\
            \textbf{Meeting Type (Status Meeting)} \\
            \quad Brainstorming Meetings & -0.47 & 0.24 & -1.94 & 0.054 & [-0.59, 0.60] \\
            \quad Customer Meetings & 0.00 & 0.30 & 0.02 & 0.987 & [-0.78, 0.20] \\
            \quad Others & -0.29 & 0.25 & -1.15 & 0.251 & [-0.19, 0.71] \\
            \quad Stand-up Meetings & 0.26 & 0.23 & 1.12 & 0.263 & [-0.70, 0.05] \\
            \textbf{Duration (Up to 30 min)} \\
            \quad 30-60 min & -0.33 & 0.19 & -1.72 & 0.088 & [-1.29, -0.14] \\
            \quad $>$60 min (*) & -0.72 & 0.29 & -2.44 & 0.016 & [-0.49, 0.43] \\
            \textbf{Day of Week (Monday)} \\
            \quad Tuesday & -0.03 & 0.24 & -0.13 & 0.898 & [-0.98, 0.08] \\
            \quad Wednesday & -0.45 & 0.27 & -1.66 & 0.099 & [-0.61, 0.40] \\
            \quad Thursday & -0.11 & 0.26 & -0.41 & 0.682 & [-0.48, 0.72] \\
            \quad Friday & 0.12 & 0.31 & 0.40 & 0.687 & [-0.61, 0.11] \\
            \textbf{Meeting Size (2-4)} \\
            \quad 5+ & -0.25 & 0.19 & -1.34 & 0.181 & [-1.13, 0.20] \\
            \textbf{Role (Listener)} \\
            \quad Discussant & -0.47 & 0.34 & -1.37 & 0.174 & [-1.44, 0.26] \\
            \quad Leader & -0.59 & 0.44 & -1.36 & 0.177 & [-0.94, 0.63] \\
            \quad Presenter/Speaker & -0.16 & 0.40 & -0.39 & 0.700 & [-0.41, 0.69] \\
            \textbf{Multitask (Low)} \\
            \quad High & 0.14 & 0.28 & 0.50 & 0.616 & [-0.52, 0.36] \\
            \textbf{Time of Day (Early Morning)} \\
            \quad Before Lunch & -0.08 & 0.22 & -0.35 & 0.730 & [-0.58, 0.89] \\
            \quad Lunch Time & 0.15 & 0.37 & 0.41 & 0.685 & [-0.62, 0.22] \\
            \quad After Lunch & -0.20 & 0.22 & -0.94 & 0.348 & [-1.17, 0.35] \\
            \quad Afternoon & -0.41 & 0.39 & -1.06 & 0.291 & [-0.44, 0.31] \\
            \textbf{Cumulative Meetings per Day (1)} \\
            \quad 2 meetings & -0.06 & 0.19 & -0.32 & 0.747 & [-0.24, 0.79] \\
            \quad $>$2 meetings & 0.28 & 0.26 & 1.05 & 0.295 & [-0.68, 1.01] \\
            \textbf{Interaction Effects} \\
            \quad Remote × Brainstorming Meetings (*) & 0.84 & 0.40 & 2.10 & 0.038 & [-1.14, 0.21] \\
            \midrule
            \multicolumn{6}{l}{$R^2$: 0.34 \quad Adj. $R^2$: 0.17 \quad F-statistic: 2.11 (p-value: 0.00)} \\
            \bottomrule
            \multicolumn{6}{l}{Significance codes: '***' 0.001 '**' 0.01 '*' 0.05} \\
        \end{tabular}
        }
        \label{tab:fixed_effects_emo}
                \vspace{-5mm}
    \end{table}

\subsection{Changes in Mutual Engagement}
\label{sec:res_mutual_eng}
To assess mutual engagement, we analyzed physiological synchrony between participant dyads using FastDTW distances. 
Because of the time-dependent nature of FastDTW, we had to normalize the measures, dividing the value by the sum of the lengths of the two time series, in line with previous work ~\cite{GiorginoJSS2009}.
Our dataset includes 46 participant pairs across various meeting durations: 25 dyads in shorter meetings (up to 30 minutes), with 22 co-located and 3 distributed pairs, while medium-length meetings (30-60 minutes) contained 21 dyads (14 co-located and 7 distributed). Although we collected data from 2 additional co-located dyads in meetings exceeding 60 minutes, we excluded these from our analysis since the absence of remote participants in this duration category prevented meaningful comparisons of physiological synchrony between participation modes.
The relative sparsity of data limited this analysis to two key predictors: participation mode (our primary focus) and meeting duration (another necessary control given the length-dependent nature of FastDTW). %measures).

The results (see Table~\ref{tab:fixed_effects_mutual}) revealed no statistically significant correlation between the main effects on physiological synchrony. 
As for participation mode, distributed dyads showed no statistically significant difference in physiological synchrony compared to co-located dyads. Similarly, for meeting duration, we found no significant difference in physiological synchrony during medium-length meetings (30-60 minutes) compared to shorter ones.
The interaction between participation mode and meeting duration was also non-significant, suggesting that the relationship between participation mode and mutual engagement remained consistent regardless of meeting length. 
%The overall model fit indicators showed that while our model captured substantial variation in mutual engagement (R$^2_c$ = 0.59), fixed effects accounted for only a small portion of this variation (R$^2_m$ = 0.06), suggesting that variability in the dyads accounts for more than meeting characteristics such as duration and participation mode.

Finally, we note that our model explained only a small portion of the variance in mutual engagement (R$^2$ = 0.06, adjusted R$^2$ = 0.01). This limited explanatory power arguably reflects the complexity of measuring mutual engagement as a phenomenon and the limitations deriving from including only two predictors due to data sparsity.
%We note that while our model captured substantial variation in mutual engagement (R$^2_c$ = 0.59), fixed effects accounted for only a small portion of this variation (R$^2_m$ = 0.06). This disparity reflects both inherent dyad variability and our necessary limitation to only two predictors (participation mode and meeting duration) due to data sparsity. Additional predictors might have increased the explanatory power of fixed effects had sufficient data been available.
Thus, we warrant caution when interpreting these findings due to the limited number of distributed dyads compared to co-located ones. %To further address potential concerns about the small sample size, particularly for distributed dyads in shorter meetings, we conducted bootstrap analyses with 500 iterations for each duration category. The resulting 95\% confidence intervals for the effect of distributed participation were [0.55, 1.56] for short meetings and [0.63, 1.38] for medium-length meetings. As both intervals include 1.0 (representing no effect), these results confirm our finding of no significant difference between participation modes.

\begin{table}[t]
    \centering
    \caption{Coefficients estimates for mutual engagement (default variables are indicated in parentheses).}
    \resizebox{\columnwidth}{!}{
    \begin{tabular}{lccccc}
        \toprule
        \textbf{Coefficients} & $\boldsymbol{\beta}$ & \textbf{SE} & \textbf{t} & \textbf{p} & \textbf{95\% CI} \\
        \midrule
        (Intercept) (***) & -2.70 & 0.16 & -16.69 & 0.00 & [-3.03, -2.38] \\
        \textbf{Dyad Participation (Co-located)} & & & & & \\
        \quad Distributed & -0.04 & 0.16 & -0.24 & 0.81 & [-0.37,\ 0.29] \\
        \textbf{Meeting Duration (Up to 30 min)} & & & & & \\
        \quad Up to 60 min & -0.29 & 0.19 & -1.53 & 0.13 & [-0.66,\ 0.09] \\
        \midrule
        \multicolumn{6}{l}{$R^2$: 0.06 \quad $Adj. R^2$: 0.01 \quad F-statistic: 1.27 (p-value: 0.29)} \\
        \bottomrule
        \multicolumn{6}{l}{Significance codes: '***' 0.001 '**' 0.01 '*' 0.05} \\
    \end{tabular}
    }
    \label{tab:fixed_effects_mutual}
            \vspace{-5mm}
\end{table}

\begin{tcolorbox}[standard jigsaw, title=Key Takeaways - Mutual Engagement, opacityback=0]%, enlarge top by=0.25cm]
    Our analysis of physiological synchrony revealed no significant differences in \textbf{mutual engagement} between co-located and distributed dyads across meetings. %The interaction effect between participation mode and meeting duration was also non-significant.
    However, the model explained minimal variance in mutual engagement due to the inclusion of only two predictors because of data sparsity. 
\end{tcolorbox}

\section{Discussion}
\label{sec:discussion}

In the following, we discuss our findings and derive empirically driven recommendations for practitioners. Furthermore, we discuss them in the context of related work on engagement in meetings, highlighting the novel contributions of the current study. Finally, we discuss the limitations to consider when interpreting our results.  

\subsection{Engagement in Hybrid Meetings}
\subsubsection*{\textbf{Main findings}}
In this study, we explored engagement in hybrid meetings by analyzing data from wearable devices and post-meeting questionnaires across three Swedish companies developing and maintaining software-intensive products and services. Our findings highlight several key insights into how hybrid meeting formats influence participants' engagement.
The evidence provided by our exploratory study reveals important insights about patterns in hybrid meetings that can help optimize collaboration practices by enhancing engagement in meetings along the three dimensions of work, emotional, and mutual engagement. 

%This evidence is observed for all dimensions of engagement in our framework. 
%Contrary to expectations, remote participants consistently reported higher work engagement levels than their onsite counterparts. However, the effectiveness of remote participation appears contingent on technological infrastructure, as remote participants with limited visibility of facial expressions and body language reported significantly lower work engagement levels.
The analysis revealed interesting role-based dynamics, with meeting listeners reporting significantly lower work engagement than other active roles (i.e., discussants and presenters). Similarly, we identified patterns related to meeting size in large meetings with 5+ participants, who consistently reported lower work engagement levels.
Moreover, afternoon scheduling corresponded with a decreased work engagement.

Notably, our empirical results mitigate concerns related to the perception of remote participants being ``second-class citizens,'' suggesting that virtual participation may not inherently lead to lower work engagement. 
However, we identified a significant interaction effect with remote participants reporting lower work engagement during medium-length meetings (30-60 minutes). 
%compared to their onsite counterparts.
This finding suggests that while remote participation does not inherently reduce engagement, extended virtual meetings may pose particular challenges for maintaining work engagement among remote participants. 
%\nicole{questo comment non discende dai findings elencati. Check}

Regarding emotional engagement, again, we found no significant differences between remote and onsite participants overall, which further challenges common assumptions about remote participants being disadvantaged. However, meeting duration emerged as a critical factor in our analysis, as we observed participants showing declining emotional engagement levels in longer meetings, regardless of participation mode.

%Additionally, our analysis revealed several other  brainstorming sessions showed lower emotional engagement compared to other meeting types, meetings held on Wednesdays exhibited lower engagement levels than other days of the week, and meetings with five or more participants demonstrated reduced emotional engagement compared to smaller meetings.
%Interestingly, while remote participation alone did not reduce engagement, we observed significant interaction effects. Remote participants actually showed higher emotional engagement during brainstorming sessions compared to their in-person counterparts, Similarly, taking an active role during the meeting (e.g. discussant) led to higher engagement levels. 
%Our analysis also identified several additional factors that corresponded with variations in emotional engagement levels. Brainstorming sessions were associated with lower emotional engagement compared to other meeting formats, whereas meetings conducted on Wednesdays showed lower engagement patterns relative to other weekdays. Similarly, meetings with five or more participants correlated with reduced emotional engagement compared to smaller ones. 

Interestingly, although remote participation by itself did not correspond with diminished engagement, we found that remote participants exhibited higher emotional engagement during brainstorming sessions compared to participants attending in person. 
This finding suggests that remote participation may coincide with enhanced emotional engagement in specific contexts. 
%In addition, participants with active roles during meetings, particularly as discussants, showed higher emotional engagement levels than those in passive roles.

%In addition, remote participants exhibited higher initial engagement in short meetings (under 30 minutes) followed by a steeper decline in medium-length meetings (30-60 minutes), whereas onsite participants maintained more consistent emotional engagement levels across different durations.

Finally, our analysis of physiological synchrony between participant dyads revealed no significant differences in mutual engagement between co-located and distributed pairs. This suggests that remote participants may achieve similar levels of mutual engagement as their onsite counterparts under appropriate conditions, though these findings should be interpreted cautiously due to sample size limitations.

\subsubsection*{\textbf{Implications for practice}} 
Our findings yield several important implications for organizations implementing hybrid work practices in software organizations. First, the comparable levels of engagement between remote and onsite participants suggest that \textit{physical co-location} may not be essential for maintaining engagement. 
Recent evidence indicates that remote meetings can actually be more effective than traditional in-person formats, being more factual and goal-oriented~\cite{lisboa2024meetings}. 
Albeit initial, our finding aligns with these observations, further challenging common assumptions about the necessity of physical presence for effective and engaging collaboration and indicating that hybrid meetings can be a viable solution when properly structured.

Concerning \textit{meeting duration}, our results suggest that all participants experienced decreased emotional engagement in meetings longer than one hour, regardless of their participation mode. This finding has direct implications for meeting planning, suggesting that shorter formats are more engaging. 
%This is particularly important for remote participants, who showed higher engagement in shorter meetings but a steeper decline in medium-length ones. 
For software organizations that have already adopted a short-meetings policy, our findings support this approach while suggesting an additional focus on maintaining high engagement through active participation; otherwise, they should consider implementing breaks or segmenting longer meetings into shorter sessions, also to maintain optimal engagement levels for both remote and onsite participants.
This result aligns with prior work reporting that short meetings, whether in-person or remote, are more effective and productive~\cite{leach2009perceived,geimer2015meetings}.

\textit{Meeting size} emerged as another crucial factor, with onsite participants in large meetings reporting significantly lower work engagement scores ($-11\%$) than remote ones. The practice of inviting participants into meetings varies across organizations based on their culture, or behavioral norms, meeting organizers should limit participation to individuals directly contributing to the meeting agenda~\cite{allen2021meeting}. This also aligns with the prior findings that a high number of meeting invitees creates confusion about the importance of attendance \cite{strayHawaii2024}. Alternative mechanisms, such as comprehensive meeting notes or decision summaries, can keep other stakeholders informed without requiring their active participation.

%\textit{Temporal factors} observed—such as lower engagement in afternoon meetings and on Wednesdays—suggest that scheduling plays a role in engagement. Organizations might consider holding meetings in the morning when the levels of energy is higher \cite{girardi_tse} and avoiding mid-week scheduling. Furthermore, attending more than two meetings per day on work engagement suggest to minimize meeting overload and encourage focused participation \cite{luong2005meetings}.

\textit{Meeting type} emerged as another important factor significantly associated with engagement. 
Our findings suggest that organizations should determine whether to allow remote participation in meetings, at least partly based on contextual factors such as the type of meeting. 
We found that remote participants showed significantly higher emotional engagement during brainstorming sessions than their in-person counterparts. This advantage may stem from greater psychological safety in sharing unconventional ideas~\cite{kark2009alive} and reduced dominance by vocal personalities when contributing from home. 
As such, organizations should consider deliberately including remote participants in brainstorming activities, providing them with meaningful opportunities to actively contribute.

Our findings also revealed  \textit{temporal patterns}—specifically, lower engagement in afternoon meetings as compared to those in the morning ($-22\%$), suggesting that meeting scheduling is significantly associated with engagement levels. Organizations might benefit from scheduling meetings in the morning hours when energy levels tend to be higher, with a positive impact on well-being and perceived productivity~\cite{girardi_tse}.
%Albeit unexpected, after interacting with managers at the three companies, we found that Wednesday is the mandatory office day, suggesting that the physical presence may increase the sense of fatigue or overload in employees who normally work remotely, potentially disrupting their routines and causing a mid-week reduction in engagement.
%Additionally, our observation that attending more than two meetings per day corresponds with lower work engagement aligns with prior research on meeting overload~\cite{luong2005meetings}, indicating that organizations should consider limiting daily meeting frequency to foster more focused participation.

%Moreover, for remote participants, the heightened engagement when acting as discussants suggests that dynamic, participatory meeting formats can mitigate some of the challenges of virtual collaboration~\cite{chen2023meetscript}. Facilitators should strive to create inclusive environments where remote participants are encouraged to contribute, leveraging their heightened engagement in these contexts. 

\textit{Active Participation} was associated with higher engagement, while passive participation (e.g., listeners) corresponded with lower engagement levels. 
%This effect was particularly evident among remote participants, for whom acting as discussants was linked to significantly higher engagement. These findings emphasize the value of active participation in fostering engagement, particularly for remote participants, who face challenges such as reduced access to social cues \cite{wainfan2004challenges} and limited opportunities for spontaneous interaction \cite{karl2022virtual, chen2023meetscript}.
Meeting facilitators should therefore work to create inclusive environments that actively encourage remote participants to contribute, such as by implementing structured turn-taking \cite{sacks1974simplest} or using interactive tools \cite{chen2023meetscript, mit_affectivu_2019}.

%The quality of \textit{video conferencing infrastructure} is another critical determinant of engagement levels. Although some studies link camera use to increased fatigue, we found work engagement levels to be significantly lower for remote participants with limited visibility of facial expressions and body language. Our findings align with previous work on agile teams working remotely \cite{reunamaki2023}. Therefore, to facilitate clear visual communication and enhance engagement, we recommend that software companies invest in adequate technology by providing high-quality video conferencing equipment and ensuring a proper setup in both corporate and home office environments~\cite{ralph2020pandemic}. Our results also highlight the importance of turning and keeping the video cameras on during meetings. Therefore, we also recommend that software organizations without an established practice or policy encourage using video in hybrid and remote meetings.  

Our findings, along with the nuanced recommendations for conducting more engaging meetings, emphasize the need for a comprehensive approach and a strong meeting culture.
This requires efforts from individual meeting participants, as well as support from the management.
While our study focused on software companies, these insights are arguably relevant for knowledge-intensive organizations across different sectors, as they face similar challenges in hybrid collaboration. 

\subsubsection*{\textbf{Ethical Considerations}}
We used physiological measurements to study engagement in hybrid meetings, which requires careful ethical consideration, in line with existing regulations on AI.\footnote{https://www.europarl.europa.eu/topics/en/article/20230601STO93804/eu-ai-act-first-regulation-on-artificial-intelligence}
We emphasize that our study does not advocate for implementing sensor-based engagement detection as a monitoring tool at the workplace. Rather, our multimodal approach to measuring engagement serves only as a methodological framework to identify factors affecting engagement in hybrid meetings, with the ultimate goal of improving meeting design practices. Furthermore, all data collection was conducted with informed consent, anonymization, and the right to withdraw.

\subsection{Related Work}
%In the following, we discuss previous works including how technology can enhance participation in hybrid meetings and how factors like meeting size and duration affect engagement. We then explain how our study builds upon and extends this previous work.

%This study fits in the vein of research on engagement in hybrid meetings. 
%In the following, we discuss previous work on engagement in hybrid meetings and highlight what are the main contributions of our work with respect to existing literature. 

%\nicole{@Daniela: We need to highlight the novelty of our contribution with respect to related work. I would structure the discussion along two aspects: methodological consideration regarding the measurement of engagement (what's new in our study? In what we replicate what others did and why) and the findings (do we align with previous findings? If not why? What are the new findings we contribute?)}

% Broad context and evolution
Research on engagement in workplace meetings has evolved from traditional in-person settings to today's hybrid environments. This evolution has driven advancements in both measurement approaches and technological solutions for detecting, monitoring, and enhancing participant engagement. 

% Evolution of measurement methods
Early work relied primarily on self-reports \cite{schaufeli2003utrecht} and observational methods. As technology advanced, researchers incorporated facial expressions \cite{monkaresi2016automated, watanabe_engauge_2023}, speech recognition \cite{ferrari2024using}, and physiological measurements \cite{dilascio2018, gashi_using, gao_n-gage_2020}. These advances in measurement techniques enabled researchers to capture different aspects of engagement, from attention and emotional states to physiological responses.
% Systems and tools enabled by measurement advances
This evolution in measurement capabilities has driven the development of engagement monitoring systems. Real-time solutions now track visual attention \cite{Akker_2009} and combine multiple data streams for comprehensive engagement assessment \cite{frank_engagement_2016}. Post-meeting analysis tools like Coco~\cite{samrose2018coco} and MeetingCoach~\cite{samrose2021meetingcoach} provide retrospective insights into engagement patterns. While these systems excel at detecting and visualizing engagement, they typically treat it as a uniform construct.
% Our contribution and approach
Our research extends beyond detection to examine engagement as a multi-dimensional phenomenon, combining physiological sensors with self-reported data to understand how different aspects of engagement manifest in hybrid settings. This approach enables us to examine engagement patterns across different participation modes and meeting characteristics.

% Key factors affecting engagement
Research has identified several critical factors affecting engagement in hybrid meetings. Meeting size consistently emerges as a significant factor, with studies by Allen et al.~\cite{allen2021meeting} and Cohen et al.~\cite{cohen2011meeting} demonstrating negative correlations between size and effectiveness. Duration similarly impacts engagement, particularly in remote settings where extended meetings increase multitasking behavior~\cite{cao2021multitasking}. 
%The visibility of social cues, especially facial expressions, plays a crucial role in maintaining engagement~\cite{kuzminykh_classification_2020, said_exploring_2020}. 
In addition, Allen et al.'s~\cite{allen2023key} observed that active participation enhances engagement in hybrid meeting discussions.

Our findings provide additional empirical evidence supporting these relationships while also revealing distinct engagement patterns between remote and co-located participants across factors such as meeting duration, size, time, type and participant roles.
% Bridge to current work
Furthermore, while prior studies have already established the importance of meeting characteristics, they have typically approached engagement as a monolithic construct. Our work advances the field by decomposing engagement into three distinct facets—work, emotional, and mutual engagement—and examining each through a novel combination of physiological measurements and self-reported data. This comprehensive approach reveals patterns in how participation mode, duration, size, and other meeting factors relate to different dimensions of engagement in hybrid settings.

\subsection{Limitations}
\label{sec:limitations}

%small dataset
Several limitations should be considered when interpreting the results, which we discuss in the following. We also explain the rationale behind our methodological choices and how we address the trade-offs associated with the decision points, in line with existing recommendations~\cite{Robillard:et:Al:2024}. 

First, the generalizability of our findings is constrained by the focus on Swedish companies, which may not represent the broader software industry, particularly given the regional variations in work culture and hybrid work adoption~\cite{R2_generalizing, wieringa2015six}.
The sample composition presents another limitation. Our dataset comprises 226 meeting participations, with an imbalance between remote (70) and onsite (156) participations, and 46 participant dyads, of which 10 were distributed and 36 were co-located. 
This imbalance, while reflecting real-world hybrid meeting compositions, may have affected our ability to detect significant differences between participation modes. Furthermore, our voluntary recruitment process may have introduced selection bias, potentially attracting participants more interested in or comfortable with hybrid meetings.
To address these limitations, we are conducting replication studies to collect larger, more balanced datasets that will strengthen the external validity of our findings. Finally, we note that, while the heterogeneous nature of our data—spanning different meeting types and organizational contexts—introduced possible confounding variables, it also enhanced the generalizability of our results across diverse organizational settings.

Methodological challenges arose from measuring engagement in a workplace setting rather than controlled laboratory conditions. We acknowledge the risk of Hawthorne effects from biometric monitoring, although participants used the devices in their natural work environments.
To reduce the risk of noise in data collection, we used comfortable, medically certified sensors\footnote{https://www.empatica.com/legal}  suitable for workplace wear, that were successfully employed in previous work~\cite{girardi_tse, Mueller:Fritz, Ferrari:etAl, Zueger:Fritz}. To further address this concern, the first author conducted daily verification of data quality to identify potential collection issues. 
While this approach sacrificed some experimental control, it enhanced ecological validity by enabling the investigation of engagement in authentic workplace contexts, which strengthens the practical implications of our findings. 
Another notable limitation was the necessary exclusion of facial expression and body language visibility variables from the analysis, despite their theoretical relevance. Their high correlation (Cramer's V = 0.88) and imbalanced distributions threatened model stability.
Although a common issue in field studies where experimental control is limited, future work might address this through targeted designs with balanced observation of non-verbal communication factors.

Finally, our physiological measurements present additional limitations. 
While electrodermal activity is validated for measuring emotional engagement, it captures only one aspect of physiological response. 
The absence of complementary measures, such as heart rate variability and facial expression analysis, limits our comprehensive understanding of engagement patterns. Future research could address this by incorporating multiple physiological measures while maintaining workplace compatibility.

\section{Conclusion}
\label{sec:conclusion}

We explored engagement patterns in hybrid meetings through a novel multi-modal approach combining physiological measurements and self-reported data. Our analysis of 226 meeting participations revealed that, contrary to common assumptions, remote participants maintained engagement levels comparable to their onsite counterparts. %However, meeting duration emerged as a critical factor, with participants in sessions exceeding 60 minutes showing decreased engagement. 
Additionally, our exploratory analysis of physiological synchrony between participant dyads suggested that remote participants achieve similar levels of mutual engagement as their onsite counterparts.

These findings have significant implications as more and more organizations are implementing hybrid work practices. The comparable engagement levels between remote and onsite participants challenge assumptions about the necessity of physical presence for effective collaboration. Our results suggest implementing shorter meeting formats, particularly for remote participants who showed lower engagement in medium-length meetings. Organizations should also avoid organizing afternoon meetings, as both remote and co-located participants reported significantly lower work engagement levels. Furthermore, meeting size should be carefully considered, as larger meetings correlate with decreased engagement among participants.

Future research could address the limitations of our study through several avenues. Expanding the participant pool to include more participants from diverse companies would enhance generalizability. Investigation of potential interventions to maintain engagement in longer meetings and enhance listener participation could provide practical solutions for current challenges. Additionally, examining the relationship between engagement patterns and meeting outcomes could offer valuable insights for optimizing hybrid collaboration. Future studies might also benefit from incorporating additional physiological measures, such as EEG, to study how aspects like fatigue and cognitive load link to engagement and participation mode.

\section*{Acknowledgements}
The research of Daniela Grassi is partially funded by D.M. 352/2022, Next Generation EU - PNRR, in the scope of the project "Recognition of emotions of cognitive workers using non-invasive biometric sensors", co-supported by Exprivia, CUP H91I22000410007.
This research was co-funded by the NRRP Initiative, Mission 4, Component 2, Investment 1.3 - Partnerships extended to universities, research centres, companies, and research D.D. MUR n. 341, 15.03.2022 – Next Generation EU (``FAIR - Future Artificial Intelligence Research", code PE00000013, CUP H97G22000210007), the Complementary National Plan PNC-I.1 - Research initiatives for innovative technologies and pathways in the health and welfare sector - D.D. 931 of 06/06/2022 (``DARE - DigitAl lifelong pRevEntion initiative", code PNC0000002, CUP B53C22006420001), and by the European Union - NextGenerationEU through the Italian Ministry of University and Research, Projects PRIN 2022 (``QualAI: Continuous Quality Improvement of AI-based Systems'', grant n. 2022B3BP5S, CUP: H53D23003510006).

\bibliographystyle{ieeetr}
\bibliography{sample} 
\balance

\end{document}